\documentclass[aps,prl,twocolumn,groupedaddress,superscriptaddress]{revtex4-1}
\usepackage{graphicx}
\usepackage{color}
\usepackage{amsmath}
\usepackage{mathtools}
\usepackage{physics}
\usepackage{epsfig}
\usepackage{wrapfig}
\usepackage{csquotes}
\usepackage{tikz}
\usetikzlibrary{shapes}
\begin{document}
\newcommand{\markercirc}{\raisebox{0.5pt}{\tikz{\node[draw,scale=0.4,circle,fill=black](){};}}}
\newcommand{\markersquare}{\raisebox{0.5pt}{\tikz{\node[draw,scale=0.4,regular polygon, regular polygon sides=4,fill=black](){};}}}
\def\CAL#1{\mathcal{#1}}	
\title{Hydroelastic wake on a thin elastic sheet floating on water}
\author{Jean-Christophe~Ono-dit-Biot}
\affiliation{Department of Physics and Astronomy, McMaster University, 1280 Main Street West, Hamilton, Ontario, L8S 4M1, Canada.} 
\author{Miguel~Trejo}
\affiliation{Laboratoire de Physico-Chimie Th\'eorique, UMR CNRS Gulliver 7083, ESPCI Paris, PSL Research University, 75005 Paris, France.}
\author{Elsie~Loukiantcheko}
\affiliation{Department of Physics and Astronomy, McMaster University, 1280 Main Street West, Hamilton, Ontario, L8S 4M1, Canada.} 
\author{Max~Lauch}
\affiliation{Department of Physics and Astronomy, McMaster University, 1280 Main Street West, Hamilton, Ontario, L8S 4M1, Canada.} 
\author{Elie~Rapha\"el}
\affiliation{Laboratoire de Physico-Chimie Th\'eorique, UMR CNRS Gulliver 7083, ESPCI Paris, PSL Research University, 75005 Paris, France.}
\author{Kari~Dalnoki-Veress}
\affiliation{Department of Physics and Astronomy, McMaster University, 1280 Main Street West, Hamilton, Ontario, L8S 4M1, Canada.} 
\affiliation{Laboratoire de Physico-Chimie Th\'eorique, UMR CNRS Gulliver 7083, ESPCI Paris, PSL Research University, 75005 Paris, France.}
\author{Thomas~Salez}
\email{thomas.salez@u-bordeaux.fr}
\affiliation{Univ. Bordeaux, CNRS, LOMA, UMR 5798, F-33405 Talence, France.}
\affiliation{Global Station for Soft Matter, Global Institution for Collaborative Research and Education, Hokkaido University, Sapporo, Hokkaido 060-0808, Japan.}
\date{\today}

\begin{abstract}
We investigate the hydroelastic wake created by a perturbation moving at constant speed along a thin elastic sheet floating at the surface of deep water. Using a high-resolution cross-correlation imaging technique, we characterize the waves as a function of the perturbation speed, for different sheet thicknesses. The general theoretical expression for the dispersion relation of hydroelastic waves includes three components: gravity, bending and stretching. The bending modulus and the tension in the sheet are independently measured. Excellent agreement is found between the experimental data and the theoretical expression.
\end{abstract}
\maketitle

In 1963, Richard Feynman described water waves in his famous \emph{Lectures on Physics}~\cite{feynman2011} as ``the worst possible example [of waves], because they are in no respect like sound and light; they have all the complications that waves can have". Several decades later, some questions remain unanswered and the study of these waves continues to be an area of great interest. For example, Kelvin's gravity wake behind a ship~\cite{kelvin1887} still stimulates fundamental questions~\cite{rabaud2013,darmon2014}. Surface tension of the liquid-air interface also influences the wave propagation, resulting in gravito-capillary waves and wake~\cite{landau1987,acheson1990,raphael1996,dias1999}. Unlike the gravity wake, the capillary wake appears ahead of the perturbation~\cite{acheson1990}. This is particularly relevant for the locomotion of insects~\cite{hu2003,chepelianskii2008,closa2010,Voise2010,steinmann2018}, as well as for nanorheological applications involving \textit{e.g.} atomic-force microscopy probes moving along thin viscous samples~\cite{wedolowski2015,ledesma2016,ledesma2017}.

Other waves of interest are the ones that propagate on elastic plates and membranes. Their properties are dictated by both the bending and stretching rigidities of the material~\cite{landau1986}. Floating such an elastic sheet on a liquid further leads to the coupling of the elastic waves to hydrodynamics. The resulting hydroelastic waves are of particular interest, as elastic sheets surrounded by fluids are ubiquitous in nature. Examples can be found in fluid mechanics~\cite{shelley2011,virot2013}, geophysics~\cite{takizawa1985,squire1988,squire1995,puaruau2002}, and biophysics~\cite{grotberg2004}. Hydroelastic waves are also relevant to practical applications in civil engineering~\cite{murai1999,watanabe2004}, as well as in energy harvesting through piezoelectric flags~\cite{akcabay2012} and control of energy radiation by trucks moving on ice sheets~\cite{davys1985}. Interestingly, the propagation of such waves can be finely controlled in an optical-like fashion by using model thin sheets with heterogeneous elastic properties~\cite{domino2018dispersion}. Different properties of these waves, such as the wave resistance or non-linear effects, have been further studied theoretically~\cite{milinazzo1995,puaruau2011}, including the overdamped limit of lubrication settings where viscosity dominates over fluid inertia~\cite{hosoi2004,vandeparre2010,lister2013,al-housseiny2013,carlson2016,arutkin2017,kodio2017}. The dispersion relation in the inertial case was analytically derived and found to depend on three components: gravity, bending and stretching~\cite{schulkes1987}. A few experimental studies developed in different contexts have studied the limiting cases where only bending and stretching~\cite{deike2013,deike2017}, or gravity and bending~\cite{takizawa1985,squire1988}, contribute. 

\begin{figure}[b!]
\includegraphics[width=1.0\columnwidth]{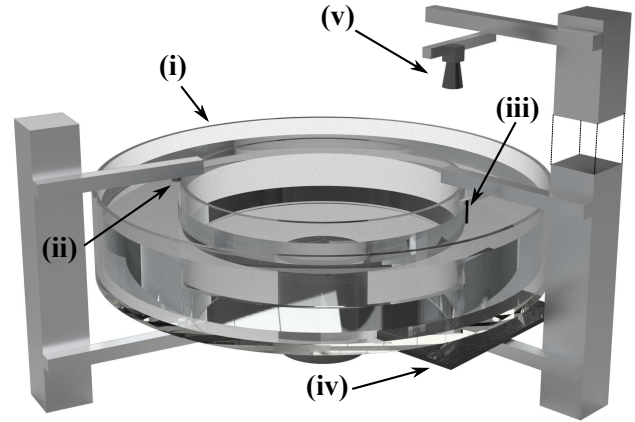}
\caption{\label{fig1} 3D schematics of the experimental setup. (i) Rotating transparent annular tank (outer radius $R_\textrm{out} = 50$~cm, inner radius $R_\textrm{in} = 30$~cm) filled with water to a depth of about 16~cm, and with an elastic sheet floating atop. (ii) Infrared beam-breaking setup to measure the angular speed $\Omega$ of the elastic sheet. (iii) A pipette perturbs the surface by blowing air, causing a wake to form. (iv) Light sheet and dot pattern used to characterize the waves using the Schlieren method~\cite{moisy2009}. (v) A camera is placed $\sim 2$~m above the tank to image the dot pattern.}
\end{figure}
In this Letter, we study the hydroelastic wake created by a perturbation moving at constant speed along an elastic sheet floating on deep water. The waves are imaged using a high-resolution optical method. By using elastic sheets with different thicknesses, the bending modulus of the sheet is varied over more than two orders of magnitude. We find excellent agreement between experimental data and the general theoretical dispersion relation, accounting for the three different contributions: gravity, bending, and stretching. 

A transparent annular tank is filled with water, as shown in Fig.~\ref{fig1}. Thin elastic sheets of Elastosil\textsuperscript{\textregistered} (Wacker Chemie AG) with nominal thicknesses $h$ of 50, 100, 200, 250 and 350~$\mu$m, and lateral dimensions of 20~cm $\times$ 16~cm are floated onto the surface of water. A thin rigid plastic support (18~cm~$\times$~1~cm~$\times$~0.1~cm) is placed atop the leading and trailing edge of the elastic sheet to ensure the sheets do not crumple. 
We experimentally verify that adding the supports does not introduce an anisotropic tension in the sheet, by ensuring that the deformation induced by ball bearings placed atop the sheet is axially symmetric (see SI). The tank is rotated at constant angular speed $\Omega$, ranging from 0 to $2.5$~rad.s$^{-1}$, causing the water to flow and the sheet to move. We take advantage of the opaque plastic supports to measure the angular speed of the sheet, using an infrared beam-breaking technique. Because of inertia, both the sheet and water do not follow the tank's speed instantaneously. Hence, all experiments are performed only once the speed of the sheet is constant and equal to the speed of the tank.

A glass capillary (World Precision Instruments, USA) is pulled to a diameter of about 100~$\mu$m at one end with a pipette puller (Narishige, Japan), and used to blow air at the surface of the sheet (Fig.~\ref{fig1}(iii)). The pipette is placed in the middle of the tank, \textit{i.e.} at a radius $R_{\textrm{p}} =40$~cm from the center. The air jet acts as a perturbation moving at speed $v= \Omega R_{\textrm{p}}$ in the reference frame of the elastic sheet, which generates an hydroelastic wake. The latter is imaged using a synthetic Schlieren method~\cite{moisy2009} involving a random dot pattern refracted by the surface topography. The dot pattern is generated using Matlab~\cite{Moisy2008,moisy2009} and printed onto a transparency film. Light shines through the dot pattern (Fig.~\ref{fig1}(iv)) and the wake, before being collected by a camera above the tank (Fig.~\ref{fig1}(v)). The Schlieren method consists in measuring the apparent displacement of the dots due to light refraction by the wake. The displacement is measured relative to a reference image of the unperturbed surface (\textit{i.e.} no air jet) moving at angular speed $\Omega$ -- which ensures that the collected information is only due to the wake. This measurement is performed using an open-source digital 2D image-correlation algorithm (Ncorr, Matlab)~\cite{blaber2015}. From the displacement of the dots, one can access the slope of the surface and thus the surface topography~\cite{moisy2009}. 

\begin{figure}[h!]
\includegraphics[width=1.0\columnwidth]{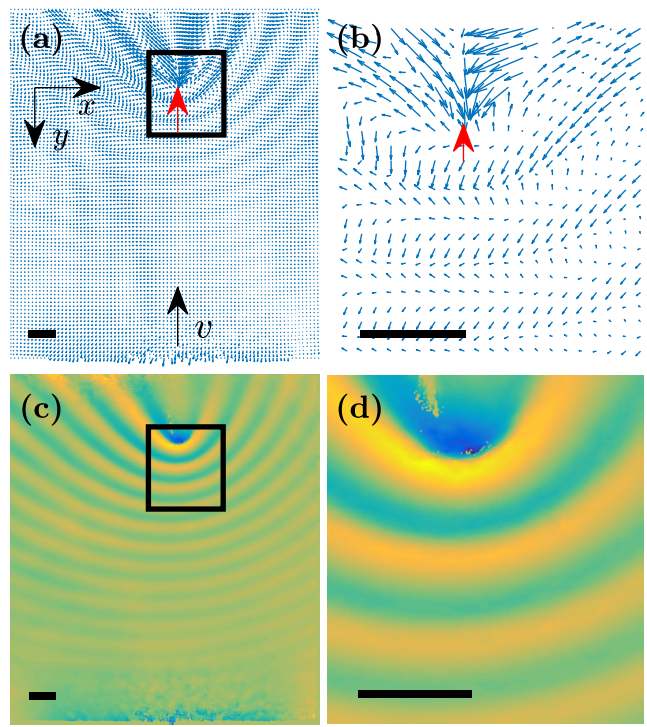}
\caption{\label{fig2} (a) Raw 2D dot-displacement data measured with Ncorr~\cite{blaber2015}, for an elastic sheet of nominal thickness $h\approx200$~$\mu$m moved at speed $v= 0.9$~m.s$^{-1}$. The displacement vectors are only shown every 10 pixels for clarity. The bottom arrow indicates the speed of the sheet, with respect to the stationary air jet whose position is indicated by the red arrow. (b) Zoom around the perturbation, corresponding to the black box in (a). (c) The $y$-component $d$ of the displacement field. Warm colours (green to orange) correspond to positive displacements, while cold colours (green to blue) correspond to negative ones. (d) Zoom around the perturbation, corresponding to the black box in (c). All scale bars correspond to 1 cm.}
\end{figure}
\begin{figure}[t!]
\includegraphics[width=1.0\columnwidth]{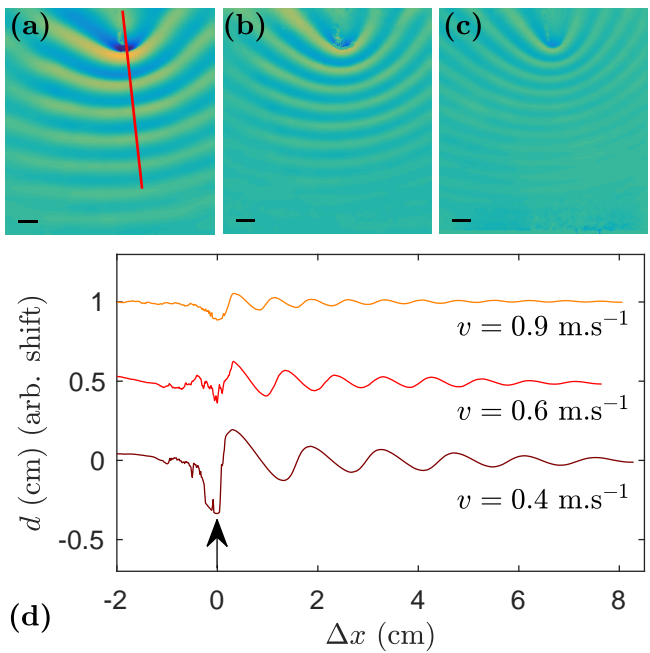}
\caption{\label{fig3} (a) - (c) $y$-component $d$ of the displacement field (see Fig.~\ref{fig2}), for speeds $v= 0.4$, 0.6, and 0.9~m.s$^{-1}$, respectively. All scale bars correspond to 1 cm. (d) $y$-component $d$ of the displacement field normal to the wave front (see red line in (a)), as a function of the distance $\Delta x$ from the perturbation, for the three speeds as indicated (shifted vertically for clarity). The position of the air-jet perturbation is indicated by the arrow.}
 \end{figure}

Figures~\ref{fig2}(a,b) show a typical vectorial displacement field. The air jet creates a localized perturbation in the sheet, as evidence by the large magnitude of the displacement field therein. Ahead of the perturbation, in the reference frame of the elastic sheet, the upstream wave pattern of the hydroelastic wake appears clearly, with a dominant -- centimetric --  wavelength $\lambda$. As the hydroelastic waves propagate along the $y$-direction, the projection of the displacement along that direction provides the strongest signal for analysis. Figures~\ref{fig2}(c,d) thus focus only on the $y$-component $d$ of the displacement field.
To characterize experimentally the dispersion relation of the hydroelastic wake, the wavelength $\lambda$ is measured as a function of the speed $v$. Figures~\ref{fig3}(a-c) show the $y$-component $d$ of the displacement field, for various speeds. The wake is slightly tilted and not symmetric about the $y$-axis, because of the centrifugal force and the surface of water assuming a parabolic profile when rotated. This distorsion is avoided in the wavelength measurement by analyzing the displacement field normal to the wave front, as shown in Fig.~\ref{fig3}(a,d). We observe both the wavelength and the displacement to decrease as the speed increases.
 
In order to quantify further and rationalize these observations, we now introduce the relevant theoretical framework. The mechanical system we consider is the thin elastic sheet in its reference frame. Neglecting the solid inertia owing to the slenderness of the sheet, the out-of-plane displacement field $z=\zeta(\boldsymbol{r},t)$ with respect to its flat horizontal state $z=0$ satisfies the F\"oppl-von K\'arm\'an equation~\cite{landau1986}:
\begin{equation}
B\nabla_{\boldsymbol{r}}^{\,4}\zeta - \sigma \nabla_{\boldsymbol{r}}^{\,2}\zeta =P + P_\textrm{ext}\ ,
\label{koppl}
\end{equation}
along the 2D horizontal space coordinate $\boldsymbol{r}=(x,y)$ and time $t$, where $\nabla_{\boldsymbol{r}}$ is the nabla operator in 2D, $B$ is the bending stiffness of the sheet, and $\sigma$ represents the tension in the sheet. The first and second terms respectively account for bending and stretching. The system is further subjected to two external forces: the excess hydrodynamic pressure $P(\boldsymbol{r},t)$ (with respect to the atmospheric one) exerted on the sheet by the water flow under gravity, and the driving pressure $P_\textrm{ext}(x,y-vt)$ modelling the perturbation by the air jet translating at constant speed $v$ along $y$. 

The water contribution, $P(\boldsymbol{r},t)$, is calculated by assuming an incompressible and irrotational flow of an inviscid fluid, in a semi-infinite half space located at $z<\zeta(\boldsymbol{r},t)$. In this context, the fluid velocity field can be written as $\nabla\varphi$, where $\varphi(\boldsymbol{r},z,t)$ is a potential that vanishes in the far field and that satisfies Laplace's equation~\cite{batchelor1967}: $\nabla^{\,2}\varphi=(\nabla_{\boldsymbol{r}}^{\,2}+\partial_z^{\,2})\varphi=0$. At lowest order in the flow (\textit{i.e.} for small-amplitude hydroelastic waves), the linearized Bernoulli equation for unsteady potential flows provides the excess hydrodynamic pressure exerted on the sheet: $P = -\rho \left. \partial{\varphi}/ \partial{t} \right|_{{z=0}}-\rho g \zeta$,  with $\rho$ the liquid density and $g$ the acceleration due to gravity. 

To obtain the dispersion relation, one substitutes the expression for $P$ into Eq.~(\ref{koppl}) in the absence of driving ($P_\textrm{ext}=0$), and invokes the kinematic condition $\left. \partial{\varphi}/\partial{z}\right|_{z=0} = \partial{\zeta}/\partial{t}$ at the water-sheet interface. Considering a plane wave $\varphi\propto\exp[i(ky-\omega t-ikz)]$ satisfying Laplace's equation, with angular wavenumber $k$ and angular frequency $\omega(k)$, yields~\cite{schulkes1987}:
\begin{equation}
\omega= \sqrt{\frac{Bk^5}{\rho}  + \frac{\sigma k^3}{\rho} + gk}\ .
\label{dispersion}
\end{equation}

\begin{figure}[h!]
\includegraphics[width=1.0\columnwidth]{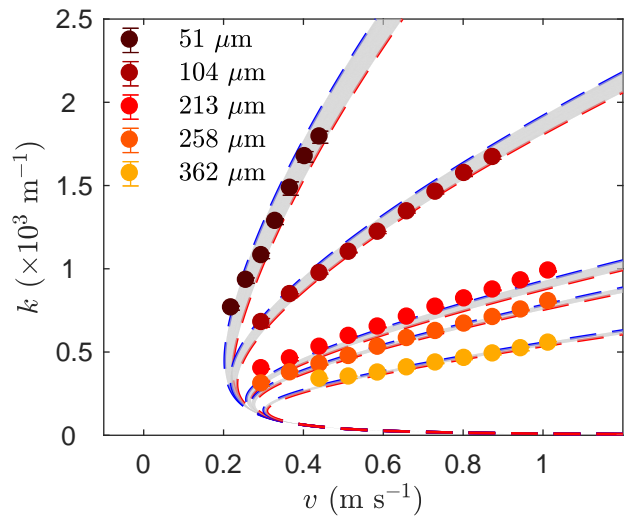}
\caption{\label{fig4} Angular wavenumber $k=2\pi/\lambda$ as a function of perturbation speed $v$, for five different sheet thicknesses $h$ as indicated in the legend. Each data point was obtained using the procedure detailed in Fig.~\ref{fig3} and an average over four different experimental displacement profiles. The error bars correspond to the standard deviation, which is comparable to the marker size. For each thickness $h$, the blue and red dashed lines separated by a grey region indicate the upper and lower theoretical predictions obtained from Eq.~(\ref{finaleq}), using the independently-measured values of $B$ and $\sigma$ (see main text) and their uncertainties.}
\end{figure}

We now consider the wake created by the driving perturbation $P_\textrm{ext}$ traveling at constant speed $v$ along $y$. In the comoving frame of the perturbation, the angular frequency $\omega'$ of a plane-wave component of the wake is shifted by the Doppler effect, and thus reads $\omega'=\omega - kv$. Furthermore, since in that comoving frame the wake is stationary, $\omega'=0$ is a necessary condition. Using Eq.~(\ref{dispersion}), one thus obtains the central relation connecting the angular wavenumber $k$ and the perturbation speed $v$, for a hydroelastic wake on deep water:
\begin{equation}
v  = \sqrt{{\frac{Bk^3}{\rho}+\frac{\sigma k}{\rho} +\frac{g}{k}}}\ .
\label{finaleq}
\end{equation}
An extensive analysis of this relation, similar to the one performed for the gravito-capillary case~\cite{raphael1996}, reveals the main features of the present wake (see also Fig.~\ref{fig4}). First, below a certain minimal speed $v^*$ wave propagation is impossible. Secondly, at a given speed $v>v^*$ there are two possible values for the observed wavelength: i) the smallest value corresponds to a group velocity that is higher than the perturbation speed $v$, and therefore the waves propagate upstream of the perturbation. This is the situation studied in the present work (see Figs.~\ref{fig2}~and~\ref{fig3}), which is  dominated by bending and stretching at sufficiently large speed; ii) the largest value corresponds to a group velocity that is lower than the perturbation speed $v$, and therefore the waves propagate downstream of the perturbation. This situation corresponds to Kelvin's classical wake~\cite{kelvin1887}, which is dominated by gravity at sufficiently large speed. 

The values of the bending modulus $B$ and the tension $\sigma$ are measured independently. The bending modulus $B = Eh^3/[12(1-\nu^2)]$ depends on three parameters: i) the Young's modulus $E = 1.11 \pm 0.06$~MPa of Elastosil\textsuperscript{\textregistered}, measured using the stress-strain curve; ii) the sheet thickness $h$, depending on the sample and measured through optical microscopy; iii) the Poisson's ratio $\nu=0.5$, assuming that Elastosil\textsuperscript{\textregistered} is an incompressible elastomer. The values of $h (\mu \mathrm m)$ and $B\mathrm{(N.m)}$ for our five different samples are found to be: 
$(h, B)= \{ 
[51\pm 1, (1.6\pm 0.1)\times10^{-8}],
[104\pm 2, (1.4\pm 0.1)\times10^{-7}],
[213\pm 7, (1.2\pm 0.1)\times10^{-6}],
[258\pm 2, (2.1\pm 0.1)\times10^{-6}],
[362\pm 3, (5.9\pm 0.3)\times10^{-6}] 
\}$. We note that the $362~\mu$m film was obtained by stacking two films with nominal thicknesses of 250~$\mu$m and 100~$\mu$m. Besides, as the sheet is freely floating on water, the tension in the sheet is equal to the air-water surface tension: $\sigma=\gamma$. The latter is measured to be $\gamma = 50 \pm 10$~mN.m$^{-1}$, as in~\cite{domino2018dispersion}, from two different techniques (see SI): i) using a Wilhelmy-plate setup; ii) characterizing the dispersion relation of gravito-capillary waves on water. The low value of $\gamma$ and the large uncertainty are attributed to the fact that the tank is filled with an important volume of tap water, and thus subject to contamination. 

As shown in Fig.~\ref{fig4} (dashed lines), using the above measured values of $B$ and $\sigma$, one can predict the evolution of the angular wavenumber $k$ as a function of the perturbation speed $v$. The uncertainties in $B$ and $\sigma$ are taken into account through two limiting predictions and an interval in between. Note that the uncertainty on $\sigma$ accounts for most of the spread between the two limiting predictions. Interestingly, for the values of $\sigma$ and $B$ considered here, all three terms in the right-hand side of Eq.~(\ref{finaleq}) are of the same order of magnitude, especially at low speed $v<0.4$~m.s$^{-1}$ (see SI). This highlights the counterintuitive role of gravity in the wavelength selection of the upstream hydroelastic wake. Finally, as expected from Eq.~(\ref{finaleq}), all the theoretical curves for different $h$ (and thus $B$) collapse onto Kelvin's gravity-dominated branch~\cite{kelvin1887}, at both large $v$ and small $k$. 

Using the experimental procedure detailed above, we measure the wavelength $\lambda$ (see Fig.~\ref{fig3}), or equivalently the angular wavenumber $k=2\pi/\lambda$, as a function of $v$. The results for the five different sheet thicknesses $h$ are shown in Fig.~\ref{fig4} (data points). We find excellent agreement between the experimental data and the theoretical predictions, with no adjustable parameter. The experimental data points for the two thinnest sheets seem to be in slightly better agreement with the upper prediction at low speed, and with the lower prediction at high speed. This observation could perhaps be related to a slight, but not quantifiable, increase in the sheet tension due to the increase in curvature of the air-water interface. Another interesting feature of Fig.~\ref{fig4} is that the difference between the upper and lower predictions decreases as the thickness $h$ of the sheet is increased. Indeed, the relative contribution of bending to Eq.~(\ref{finaleq}) increases, and the difference between both predictions, which is mainly due to the uncertainty in tension, decreases. Note that Kelvin's classical gravity-dominated branch~\cite{kelvin1887} corresponds to: i) a wake propagating behind the perturbation; ii) a wavelength that would almost reach the meter range in our experiments, which is not attainable with the current setup. 

In this Letter, we have studied the hydroelastic wake formed by moving a thin elastic sheet, floating on water, past a stationary air jet. Specifically, we experimentally measured the wavelength of the wake as a function of the perturbation speed, for sheets with bending moduli varying over two orders of magnitude. For thin elastic sheets (thickness smaller than $100$~$\mu$m), stretching plays a significant role in the propagation of the waves. For thicker elastic sheets (thickness larger than $100$~$\mu$m), the bending contribution becomes dominant -- a regime that is particularly relevant for floating ice~\cite{takizawa1985,squire1988,davys1985}. The results are found to be in excellent agreement with theoretical predictions, based on the elasticity of slender structures coupled to the hydrodynamics of inviscid incompressible flows, with no adjustable parameter. Interestingly, for thin elastic sheets, bending, stretching and gravity all contribute to the hydroelastic wake -- a result with practical consequences in geophysics, biophysics and civil engineering.

The financial support by the Natural Science and Engineering Research Council of Canada and the Joliot chair of ESPCI Paris is gratefully acknowledged. The authors thank the Global Station for Soft Matter, a project of Global Institution for Collaborative Research and Education at Hokkaido University. They are also grateful to Andreas Koellnberger and Wacker Chemie AG for technical information and the donation of Elastosil\textsuperscript{\textregistered} films, as well as to Antonin Eddi, Lucie Domino, Andreas Carlson and Yacine Amarouchene for stimulating discussions. 
\bibliography{Onoditbiot2018.bib}
\end{document}


\title{Supplemental Information for:\\ ``Hydroelastic wake on a thin elastic sheet floating on water''} 
\author{Jean-Christophe~Ono-dit-Biot}
\affiliation{Department of Physics and Astronomy, McMaster University, 1280 Main Street West, Hamilton, Ontario, L8S 4M1, Canada.} 
\author{Miguel~Trejo}
\affiliation{Laboratoire de Physico-Chimie Th\'eorique, UMR CNRS Gulliver 7083, ESPCI Paris, PSL Research University, 75005 Paris, France.}
\author{Elsie~Loukiantcheko}
\affiliation{Department of Physics and Astronomy, McMaster University, 1280 Main Street West, Hamilton, Ontario, L8S 4M1, Canada.} 
\author{Max~Lauch}
\affiliation{Department of Physics and Astronomy, McMaster University, 1280 Main Street West, Hamilton, Ontario, L8S 4M1, Canada.} 
\author{Elie~Rapha\"el}
\affiliation{Laboratoire de Physico-Chimie Th\'eorique, UMR CNRS Gulliver 7083, ESPCI Paris, PSL Research University, 75005 Paris, France.}
\author{Kari~Dalnoki-Veress}
\affiliation{Department of Physics and Astronomy, McMaster University, 1280 Main Street West, Hamilton, Ontario, L8S 4M1, Canada.} 
\affiliation{Laboratoire de Physico-Chimie Th\'eorique, UMR CNRS Gulliver 7083, ESPCI Paris, PSL Research University, 75005 Paris, France.}
\author{Thomas~Salez}
\email{thomas.salez@u-bordeaux.fr}
\affiliation{Univ. Bordeaux, CNRS, LOMA, UMR 5798, F-33405 Talence, France.}
\affiliation{Global Station for Soft Matter, Global Institution for Collaborative Research and Education, Hokkaido University, Sapporo, Japan.}
\date{\today}
\begin{abstract}
\end{abstract}
\pacs{}
\renewcommand{\thefigure}{S\arabic{figure}}
\renewcommand{\theequation}{S\arabic{equation}}
\renewcommand{\thetable}{S\arabic{table}}
\maketitle

\section{Tension isotropy}
\begin{figure}[h!]
\includegraphics[width=13cm]{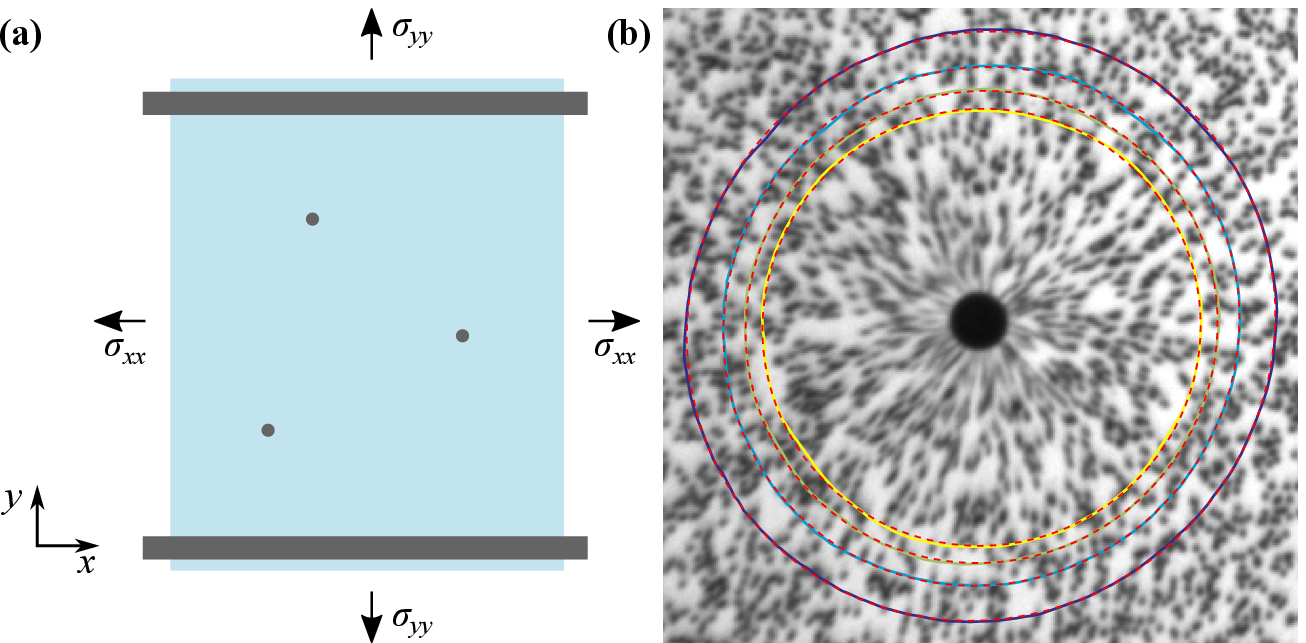}
\caption{\label{tension} (a) Top-view schematic of the elastic sheet floating on water. Plastic beams are placed at the leading and trailing edges of the elastic sheet to ensure the sheet does not crumple. Ball bearings are dropped atop the sheet to verify that no anisotropy is introduced when placing the beams. (b) Picture of the reference dot pattern, seen through the water and the sheet, around a ball bearing. The solid lines are sample isodisplacement lines. The dashed lines are the best fits of the isodisplacement lines to ellipses. The best-fit ellipticities are found to be equal to 1 for all cases.}
\end{figure}
A schematic of the elastic sheet is shown in Fig.~\ref{tension}(a). The tensions along the $x$- and $y$-axes are denoted $\sigma_{xx}$ and $\sigma_{yy}$, respectively. We place ball bearings directly on the sheet floating on water, and we image the resulting deformation using the optical Schlieren method~\cite{moisy2009}. We then calculate the magnitude of the displacement vector field, which is directly linked to the deformation of the elastic sheet. Sample isodisplacement lines are shown in Fig.~\ref{tension}(b). We quantify the anisotropy of the deformation by fitting the isodisplacement lines to ellipses. The best-fit ellipticities for the four cases shown in Fig.~\ref{tension}(b) are all found to be equal to 1, meaning that the isodisplacement lines are circles, which thus indicates that the tension in the sheet is isotropic. Indeed, if $\sigma_{xx}$ or $\sigma_{yy}$ was larger than the other, the deformation would be elongated along the low-tension direction, leading to an ellipticity larger than 1. 

\section{Tension measurement}
The tension $\sigma_{xx}$ along the $x$-axis is set by the water-air surface tension $\gamma$, as both the left and right edges are free (see Fig.~\ref{tension}(a)). Since all the experiments presented in the study are performed on sheets where the tension is isotropic, one can safely assume that $\sigma=\sigma_{yy}=\sigma_{xx}=\gamma$. The tabulated value for the pure water-air surface tension under ambient conditions is $\gamma=72$~mN.m$^{-1}$, but it is extremely sensitive to contamination by all kinds of surfactants. The experiments being conducted in an open tank containing $\sim 200$~L of water, contamination is unavoidable. Therefore, $\gamma$ was measured from two independent methods. 

First, using a Wilhelmy-plate setup, $\gamma$ was found to be between $40$ and $55$~mN.m$^{-1}$, for water from three different sources: water from the tank after one day, tap water, and deionized water. The largest value of $\gamma$ was obtained for deionized water, and the smallest one for the water from the tank -- which is consistent with tank contamination over time.
\begin{figure}[h!]
\includegraphics[width=10cm]{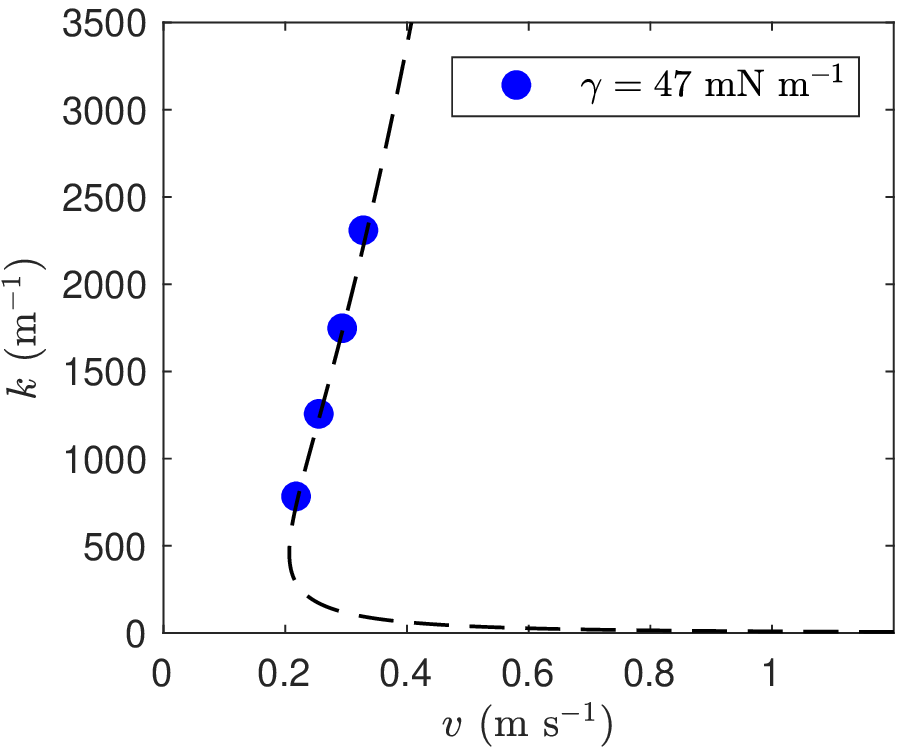}
\caption{\label{capillary} Angular wavenumber $k$ as a function of perturbation speed $v$, for the gravito-capillary wake at the surface of deep water. The dashed line shows the best fit of Eq.~(\ref{kvsv}) to the experimental data points. The water-air surface tension $\gamma = 47$~mN.m$^{-1}$ is obtained as the only adjustable parameter.}
\end{figure}

Another approach to measure $\gamma$ is to invoke the gravito-capillary wake formed at the surface of deep water by a perturbation moving at constant speed $v$. In such a case, the analogue of Eq.~(3) is~\cite{raphael1996}:
\begin{equation}
v  = \sqrt{{\frac{\gamma k}{\rho} +\frac{g}{k}}}\ .
\label{kvsv}
\end{equation}
Therefore, $\gamma$ can be evaluated by fitting Eq.~(\ref{kvsv}) to the experimental evolution of the angular wavenumber $k$ as a function of speed $v$ for a gravito-capillary wake. In fact, as the elastic sheet only covers a small portion of the water in the tank, the hydroelastic wake is only observed once a lap when the sheet moves across the stationary perturbation at speed $v$. Otherwise, water flowing at speed $v$ is directly exposed to the perturbation, and a gravito-capillary wake is instead formed at the surface. Using the Schlieren method, the wavelength $\lambda=2\pi/k$ of the upstream gravito-capillary wake is measured as a function of speed $v$. The measurements of the wavelengths for both the hydroelastic and the gravito-capillary wakes are thus performed simultaneously. Figure~\ref{capillary} shows the evolution of the angular wavenumber $k$ as a function the speed $v$, for the gravito-capillary wake. By fitting the experimental data to Eq.~(\ref{kvsv}), one finds $\gamma = 47$~mN.m$^{-1}$. 

Note that the last measurement was performed during the characterization of the hydroelastic wake on a sheet of thickness $h=50$~$\mu$m. Similar measurements were also performed during the characterization of the hydroelastic wake on sheets with larger thicknesses $h = \{100; 200; 250\}$~$\mu$m. However, in those cases, the wavelength for the gravito-capillary wake was measurable only for the lowest speed, $v\approx0.2$~m.s$^{-1}$.  For all those three measurements, we get $\gamma = 47$~mN.m$^{-1}$. Considering all the measured values from both methods, we reach the conclusion that $\gamma = 50 \pm 10$~mN.m$^{-1}$. 

\section{Contributions of bending, stretching and gravity}
\begin{figure}[h!]
\includegraphics[width=15cm]{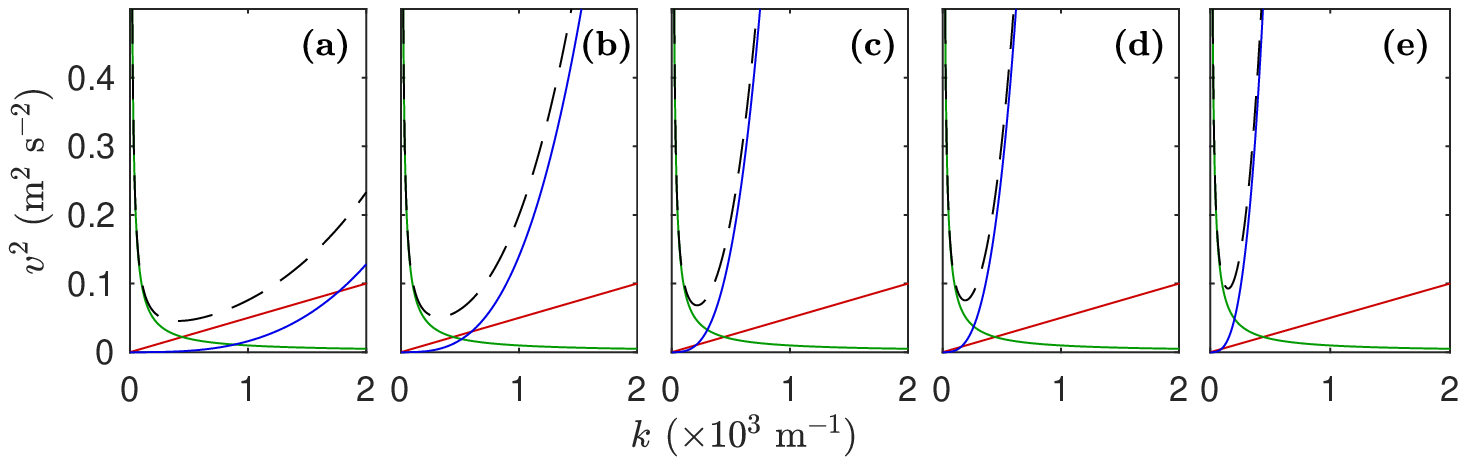}
\caption{\label{contributions} Contributions of bending, stretching and gravity in Eq.~(3). The five  panels correspond to different thicknesses (and thus bending moduli $B$) -- from left to right: $h = 50, 100, 200, 250$, and 350~$\mu$m. The blue line corresponds to bending ($B k^3/\rho$), the red line to stretching ($\sigma k/\rho$), and the green line to gravity ($g/k$); while the dashed line represents the sum of the three contributions.}
\end{figure}

Let us consider Eq.~(3). The first term in the square root corresponds to bending, the second one to stretching, and the third one to gravity. Using all the measured values for $B$, and the value of $\sigma$, the respective contributions of those three terms, as well as their sum, are computed from Eq.~(3) and plotted in Fig.~\ref{contributions}. We experimentally measure angular wavenumbers ranging from 400 to 2000~m$^{-1}$ (see Fig.~4). For the thinnest film $h=50$~$\mu$m (Fig.~\ref{contributions}(a)), all three terms do contribute in that range. The elastic sheet with $h=100$~$\mu$m shows an interesting behaviour (Fig.~\ref{contributions}(b)): at low angular wavenumbers ($k<1000$~m$^{-1}$), all three terms are relevant, while bending becomes predominant at larger angular wavenumbers. Finally, for the three largest thicknesses, $h = 200, 250$, and 350~$\mu$m ((Fig.~\ref{contributions}(c-e))), bending clearly dominates.
\bibliography{Onoditbiot2018_SI.bib}